\documentclass[conference]{IEEEtran}
\IEEEoverridecommandlockouts
\usepackage{cite}
\usepackage{amsmath,amssymb,amsfonts}
\usepackage{algorithmic}
\usepackage{graphicx}
\usepackage{textcomp}
\usepackage[multiple]{footmisc}
\usepackage{multirow}
\usepackage{listings}
\usepackage{subfig}
\usepackage{makecell}
\usepackage{balance}
\usepackage[table,xcdraw]{xcolor}
\usepackage{eso-pic}
\def\BibTeX{{\rm B\kern-.05em{\sc i\kern-.025em b}\kern-.08em
T\kern-.1667em\lower.7ex\hbox{E}\kern-.125emX}}

\usepackage[pdftex]{hyperref}
\usepackage{tikz}

\newcommand\copyrighttext{%
    \footnotesize \textcopyright 2022 IEEE. Personal use of this material is permitted.
    Permission from IEEE must be obtained for all other uses, in any current or future
    media, including reprinting/republishing this material for advertising or promotional
    purposes, creating new collective works, for resale or redistribution to servers or
    lists, or reuse of any copyrighted component of this work in other works.
    DOI: \href{https://doi.org/10.1109/IPCCC55026.2022.9894299}{https://doi.org/10.1109/IPCCC55026.2022.9894299}}
\newcommand\copyrightnotice{%
    \begin{tikzpicture}[remember picture,overlay]
        \node[anchor=south,yshift=10pt] at (current page.south) {\fbox{\parbox{\dimexpr\textwidth-\fboxsep-\fboxrule\relax}{\copyrighttext}}};
    \end{tikzpicture}%
}

\begin{document}

    \title{Reshi: Recommending Resources for Scientific Workflow Tasks on Heterogeneous Infrastructures\\
    }

    \author{

        \IEEEauthorblockN{Jonathan Bader\IEEEauthorrefmark{1}, Fabian Lehmann\IEEEauthorrefmark{3}, Alexander Groth\IEEEauthorrefmark{1}, Lauritz Thamsen\IEEEauthorrefmark{4},  \\
        Dominik Scheinert\IEEEauthorrefmark{1},
        Jonathan Will\IEEEauthorrefmark{1},  Ulf Leser\IEEEauthorrefmark{3}, Odej Kao\IEEEauthorrefmark{1}}

        \IEEEauthorblockA{
            \IEEEauthorrefmark{1}
            \{jonathan.bader, a.groth, dominik.scheinert, will,  odej.kao\}@tu-berlin.de,
            Technische Universität Berlin, Germany
        }
        \IEEEauthorblockA{
            \IEEEauthorrefmark{3}
            \{fabian.lehmann, leser\}@informatik.hu-berlin.de,
            Humboldt-Universität zu Berlin, Germany
        }
        \IEEEauthorblockA{
            \IEEEauthorrefmark{4}
            lauritz.thamsen@glasgow.ac.uk,
            University of Glasgow, United Kingdom
        }
  
    }

    \maketitle
    \copyrightnotice
    \pagestyle{plain}

    \begin{abstract}
        Scientific workflows typically comprise a multitude of different processing steps which often are executed in parallel on different partitions of the input data. These executions, in turn, must be scheduled on the compute nodes of the computational infrastructure at hand. This assignment is complicated by the facts that (a) tasks typically have highly heterogeneous resource requirements and (b) in many infrastructures, compute nodes offer highly heterogeneous resources. In consequence, predictions of the runtime of a given task on a given node, as required by many scheduling algorithms, are often rather imprecise, which can lead to sub-optimal scheduling decisions.

We propose Reshi, a method for recommending task-node assignments during workflow execution that can cope with heterogeneous tasks and heterogeneous nodes. Reshi approaches the problem as a regression task, where task-node pairs are modeled as feature vectors over the results of dedicated micro benchmarks and past task executions. Based on these features, Reshi trains a regression tree model to rank and recommend nodes for each ready-to-run task, which can be used as input to a scheduler. For our evaluation, we benchmarked 27 AWS machine types using three representative workflows. We compare Reshi's recommendations with three state-of-the-art schedulers.
Our evaluation shows that Reshi outperforms HEFT by a mean makespan reduction of 7.18\% and 18.01\% assuming a mean task runtime prediction error of 15\%. 
    \end{abstract}


    \section{Introduction}\label{sec:INTRO}
    The popularity of scientific workflows increases and has become essential in many scientific domains like bioinformatics, geoinformatics, or physics~\cite{ewels2020nf,da2020characterizing,nextflow, deelman2019evolution, mohammadi2019integer, bader2021tarema,lehmann2021force,baderLotaruLocallyEstimating2022}.
For example, as new genome sequencing machines, satellites, and microscopes produce more and fine-granular data, workflows become more complex and have to deal with larger datasets.
Accordingly, the execution of a single workflow can take multiple days or even weeks on big clusters~\cite{da2020characterizing, bader2021tarema, baderLotaruLocallyEstimating2022}.

Scientific workflows consist of a set of black box tasks and a set of directed edges which describe the dependencies between the tasks.
Accordingly, a predecessor task has to finish before the successor task can start. 
The tasks communicate via files and the output of a predecessor task is the input of its successor. 
Workflows following this definition can be represented as a Directed Acyclic Graph (DAG).

As workflows can comprise thousands of black box task instances, scientific workflow management systems (SWMS) like  Pegasus~\cite{deelman2019evolution} or Nextflow~\cite{nextflow} are used to reduce a scientist’s manual configuration effort by, for example, execution monitoring, automatic parallelization, or improved reusability.
With the help of resource managers like Kubernetes~\cite{kubernetes} or Slurm~\cite{slurm}, the SWMS schedule the tasks to the available cluster nodes.

Some clusters support multiple purposes and, therefore, consist of heterogeneous nodes. Other clusters get upgraded over time or get partially replaced with newer components once hardware failures occur, leading to a heterogeneous infrastructure~\cite{hutson2019managing,will2021c3o, kulagina2022hetpart, bader2021tarema, baderLotaruLocallyEstimating2022}.
However, even nodes with the same amount of CPU cores, memory, or disk space can yield highly different runtimes, for example when memory latencies, clock speeds, or read/write rates differ.
Furthermore, different tasks often have different resource demands.
For instance, some tasks are very CPU-heavy or memory-intensive, while others mostly read and write to the disk~\cite{da2020characterizing, da2015online, bader2021tarema}.
Hence, cluster resource heterogeneity can be exploited by schedulers for optimized task placements.

Despite the extensive research on scheduling algorithms that incorporate heterogeneity aspects, there is a lack of use in existing real-world systems.
The missing uptake can be explained with a lack of accurate task runtime estimates since even state-of-the-art estimators yield a median prediction error between 10\% and 20\%~\cite{da2015online, nadeem2017modeling, pham2017predicting, hilman2018task, baderLotaruLocallyEstimating2022}.
In practice, popular resource managers handle tasks as black boxes and, therefore, often resort to simpler scheduling approaches like Round-robin or fair scheduling~\cite{bader2021tarema, baderLotaruLocallyEstimating2022,carrion2022kubernetes, tang2016fair}.
For example, Kubernetes uses a Round-robin strategy to assign tasks to nodes~\cite{carrion2022kubernetes}, while YARN does it in a fair fashion~\cite{tang2016fair}.

In this paper, we propose a recommender system for heterogeneous infrastructures (Reshi) to rank machines for scientific workflow tasks without being affected by error-prone runtime estimates.
Our approach consists of four steps.
During the first step, we profile the existing cluster hardware with a set of benchmarks to gather detailed performance insights.
Then, Reshi analyzes existing task performance metrics from historical workflow executions.
In case of no existing executions a quick workflow profiling with highly reduced inputs can be conducted to gather task execution metrics~\cite{baderLotaruLocallyEstimating2022, quasar}.
Based on the infrastructure details and the task-performance data, we train a regression tree model for our recommender system to dynamically rank the nodes for each task.
Through these ranks, we avoid relying on accurate runtime estimates and data-dependent resource usage predictions.
The ranks can then be used to schedule workflow tasks without assuming accurate runtime knowledge.

We provide a practical implementation of our approach as a prototype, comprising a cluster benchmarking tool for heterogeneous clusters and the recommender system for ranking the tasks to the infrastructure.
For our evaluation, we benchmarked and profiled 27 different AWS EC2 machines to evaluate our prototype using three real-world scientific workflows from the popular nf-core framework~\cite{nextflow}. 
Further, we provide a WorkflowSim extension\footnote{github.com/CRC-FONDA/WorkSim-PredError} that enables the lookup of actually measured runtimes and the inclusion of prediction errors in the simulation.
Based on this simulation extension, our comparison with the state-of-the-art schedulers HEFT, MinMin, and Round-robin shows that Reshi helps the SWMS to achieve low workflow makespans while being independent from task prediction errors.



    \section{Related Work}\label{sec:RELATED_WORK}
    In this section, we first cover the scheduling of scientific workflows on heterogeneous clusters.
Then, we focus on runtime prediction approaches since their task runtime estimates frequently serve as the input for related scheduling approaches~\cite{baderLotaruLocallyEstimating2022}.

\subsection{Scheduling Scientific Workflows on Heterogeneous Clusters}

Existing approaches either consider workflow scheduling in a statically or dynamically manner~\cite{taxonomyscheduling}.
Static scheduling heuristics like HEFT~\cite{heft} or HCPPEFT~\cite{dai2014synthesized} assign a set of tasks to compute resources before workflow execution.
Dynamic approaches like P-HEFT~\cite{pheft} map tasks to infrastructure components at runtime or are able to adjust their scheduling plan dynamically.
Therefore, these techniques are more flexible to changes in the actual execution, e.g., node failures.
However, the presented state-of-the-art scheduling approaches have in common that they need extensive knowledge about the physical DAG, execution times on all machines, and communication times between dependent tasks.
Accordingly, these approaches are often not feasible in real-world systems due to the comprehensive knowledge that is required.

In our approach, the employed and trained recommender system ranks machines for task instances on demand.
These ranks can then be used by a more sophisticated scheduler to reduce the dependence from task runtimes, e.g., a scheduling approach that works with ranks and avoids the usage of error-prone task runtime estimates.

\subsection{Task Runtime Prediction}
Some approaches predict the runtime of tasks in a workflow~\cite{da2015online, nadeem2017modeling, baderLotaruLocallyEstimating2022, hilman2018task, pham2017predicting}.
The papers employ several prediction models to find a reasonable runtime estimate. 
Therefore, they use regression, decision trees, neural networks, or other statistical approaches to create prediction models.
As we use the runtime estimation approaches as a reference for state-of-the-art task runtime prediction errors, we will describe them in more detail in Section~\ref{subsec:workSim}.

Our approach explicitly avoids predicting runtimes or relying on runtime predictions.
Instead, our aim is to rank machines for a task, which can work as a substitute for the exact runtime.
Therefore, different input parameters, like larger datasets, would influence the runtime but not necessarily the rank.

    \section{Reshi Approach}\label{sec:APPROACH}
\begin{figure}
    \includegraphics[width=\columnwidth]{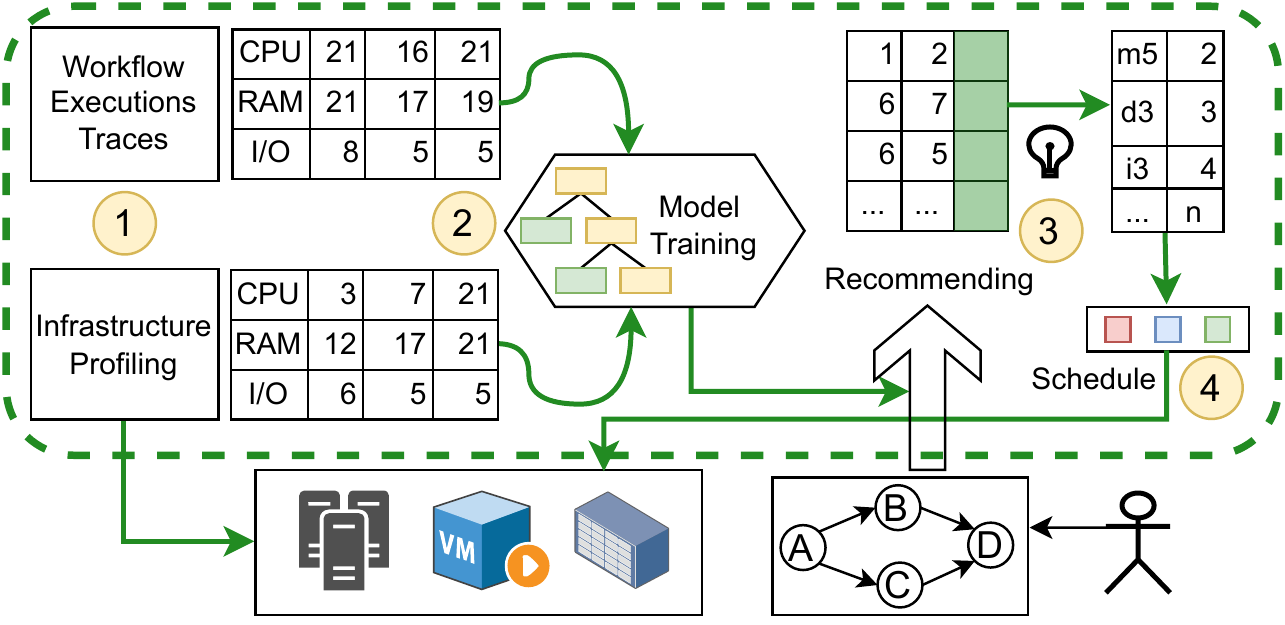}
    \caption{Our approach in green with step \textcircled{\raisebox{-0.9pt}{1}} to \textcircled{\raisebox{-0.9pt}{4}}.} \label{fig:arch}
\end{figure}

    This section gives an overview of our proposed system. 
The numbering matches the circled numbers in Fig. \ref{fig:arch}. \\

In the \textcircled{\raisebox{-0.9pt}{1}} \textbf{Cluster Profiling} step, we profile the available nodes in the cluster with a test suite that contains several benchmarks.
Additionally, we add and align available data from historical workflow executions and the contained tasks to the data repository.
If there are no existing workflow executions, an additional short workflow execution with highly reduced data inputs can be conducted to gather initial traces~\cite{baderLotaruLocallyEstimating2022}.
In the \textcircled{\raisebox{-0.9pt}{2}} \textbf{Model Creation} step, we create a regression tree model based on the previously gathered infrastructure attributes like CPU speed, memory speed, or I/O.
We enrich the model with task monitoring data from the historical workflow executions.
This model is intended to rank and recommend nodes for a certain task in a heterogeneous cluster.
The ranking is done in the \textcircled{\raisebox{-0.9pt}{3}} \textbf{Ranking and Recommendation} step.
Thereby the model uses one submitted task as the input and then creates a ranking of fitting nodes.
This ranking is used in the \textcircled{\raisebox{-0.9pt}{4}} \textbf{Scheduling} step, serving as a recommendation for the scheduler to place a task in the cluster.

\subsection*{\textcircled{\raisebox{-0.9pt}{1}} Cluster Profiling}
\label{subsec:profiling}

We assume the compute cluster consists of several nodes with different kinds of hardware.
Initially, we gather performance characteristics.
For this, we use various benchmarks to measure CPU, memory, and disk I/O characteristics.
Benchmarking the heterogeneous cluster can be done in parallel and has to be conducted only once for each node.
Once failures or changes are detected in the cluster, the cluster resource manager, e.g. Kubernetes or Slurm, can rerun the profiling on the changed nodes.
Instead of working with the absolute measurement values, we rank the nodes for each benchmarked feature.

The task performance data originates from historical workflow executions or workflow profiling that uses the monitoring part of the SWMS.
The data contains CPU, memory, disk I/O characteristics.

\subsection*{\textcircled{\raisebox{-0.9pt}{2}} Model Creation}
\label{subsec:model-creation}

The recommender system uses a subset $C_{exc}$ of all possible task-node combinations $C$ of task executions on the possibly heterogeneous infrastructure, i.e., $C_{exc} \subseteq C$.
This input enables Reshi to learn an effective mapping and allow for recommendations of promising combinations.
Based on user-defined conditions, e.g., a change in the cluster architecture, the recommender system can be scheduled for retraining.

In a next step, the data obtained from profiling the $i$-th cluster node and executing the $j$-th task is consolidated into an input vector $\vec{c}_{ij}\in \mathbb{R}^{v+w}$, where $v$ denotes the number of task-related metrics and $w$ the number of resource-related metrics respectively.
We hereby obtain a vector for each executed combination entry $c_{ij}\in C_{exc}$. 
The gathered information is then transformed into a matrix $X^{n \times l}$, with $n = |C_{exc}|$ and $l = v+w$.
We use this matrix as a training input for our regression tree model.

\subsection*{\textcircled{\raisebox{-0.9pt}{3}} Ranking and Recommendation}
\label{subsec:ranking-recommendation}

Each node $m_i$ in the set of nodes $M$ has $q \in Q$ available resources, e.g. CPU, memory, I/O, GPU, denoted as $rm_i^{(q)}$.
At the same time, each task $t_j$ in the set of tasks $T$ formulates requirements with respect to the different resources, denoted as $rt_j^{(q)}$. 
Therefore, our goal is to establish a ranking of nodes and to select the best fitting one for each given task.

First, we filter the set of nodes $M$ to select all nodes that fulfill the task's resource requirements.
We define the set of allocatable nodes as 
\begin{equation*}
M_{alloc} = \{ m_i \in M | \forall q \in Q: rm_i^{(q)} \geq rt_j^{(q)}\},
\end{equation*}
i.e., in order to be considered as allocatable, the nodes need to have enough available resources to fulfill the resource request of the respective task $t_j$.
Then, we pass the physical task $t_j$ together with the set of allocatable nodes $M_{alloc}$ to our regression tree model.
The model then ranks the task $t_j$ for each node $m_i \in M_{alloc}$ and creates a node priority list $P$ in ascending order according to the ranking.
The node with the lowest list index is the most recommended node.
However, the priority list $P$ can be used by the scheduling unit to, for example, optimize over a list of available tasks inside the queue and their ranks.

\subsection*{\textcircled{\raisebox{-0.9pt}{4}} Usage in Scheduling}
\label{subsec:appraoch-sched}

In the last step, a scheduler has to assign the tasks to the best fitting node.
We propose two simple prioritizing approaches that use Reshi's recommendations.

The first task prioritizing strategy, Reshi-C, compares the number of children tasks and prefers tasks with many children.
The second strategy, Reshi-M, orders all tasks by the average runtime from historical executions descending.
Both strategies then allocate nodes to the tasks using the recommendations and the ordered queue.

These prioritizing strategies serve as simple examples as we intend to show how the ranking could be used.
More sophisticated schedulers could substitute their dependence from runtimes per task-node pair through our provided ranking.

    \section{Experiment Setup}\label{sec:EXPERIMENTALSETUP}
    In this section, we describe profiling benchmarks for heterogeneous clusters, our WorkflowSim~\cite{chen2012workflowsim} extension that can incorporate prediction errors, the evaluation workflows, and the baselines.

\subsection{Prototype Implementation}
\label{subsec:prototype-implementation}
This subsection shortly explains the infrastructure profiler and the recommender implementation.
\paragraph{Infrastructure Profiler}
For profiling and benchmarking, we build on the Phoronix Test Suite\footnote{phoronix-test-suite.com, Accessed: June 2022}.
The Phoronix Test Suite supports the installation, execution, and monitoring of a large variety of benchmarks.

To determine the CPU's maximum throughput of hashes per second, we use \textit{John the Ripper} (JtR)\footnote{github.com/openwall/john, Accessed: June 2022}.
Additionally, as a second metric of CPU performance, we use the time required to build the Linux kernel (BLK)\footnote{kernel.org, Accessed: June 2022}. 
While JtR is fully CPU-bound, BLK is largely CPU bound but can be impacted by the I/O for extremely low build times. 

For memory performance measurements, we use \textit{RAMspeed}\footnote{alasir.com/software/ramspeed, Accessed: June 2022} and \textit{Stream}\footnote{cs.virginia.edu/stream, Accessed: June 2022}. 
Both tools run four different operations, namely COPY, SCALE, ADD, and TRIAD.
We combine both tools for higher accuracy in measuring the performance of the RAM.

Lastly, \textit{fio} measures the data transfer rate and IOPS of the storage medium for sequential and random access.
We chose block sizes of 4KB and 2MB for the random and sequential tests, respectively. 

\paragraph{Recommender Implementation} 
For our recommender implementation, we use a regression tree model. 
The profiling values from the nodes, CPU, memory, and I/O metrics, serve as the first part of the input vector.
The second part of the input vector contains the task traces, for example, the CPU usage, read/written bytes, peak memory, and average memory usage.

\subsection{Workflow Simulation}
\label{subsec:workSim}

Established simulation tools, like WorkflowSim~\cite{chen2012workflowsim} or WRENCH~\cite{casanova2018wrench} assume accurate task runtime knowledge.
However, task runtime predictions inherently yield a certain prediction error.
We want to incorporate such a systematic prediction error into our simulations.

Further, they define a certain number of MIPS (millions of instructions per second) to a node in the cluster and use this value to calculate the runtime depending to the node.
However, this oversimplification neglects that tasks show different resource access patterns, e.g., particular applications run faster or slower on different CPUs architectures, while other tasks are solely I/O bound. 

To overcome these limitations, we extend the popular WorkflowSim simulation environment in our simulation setup.

First, to incorporate the systematic task runtime prediction error, we evaluated the papers from Section~\ref{sec:RELATED_WORK} in order to derive realistic prediction errors.
Nadeem et al.~\cite{nadeem2017modeling} report a normalized average absolute prediction error of 10\%, 11\%, and 15\% for three different workflows.
For the tasks in two workflows, the error is normally distributed, while the third workflow yields that the majority of tasks show higher error rates.
Hilman~\cite{hilman2018task} report the prediction errors for all tasks in a single workflow.
They show that their technique is able to provide a task runtime estimation error below 5\% for two tasks but also errors above 30\% for three other workflow tasks.
They are able to outperform the Two-stages baseline~\cite{pham2017predicting}, where the authors report a slightly higher relative absolute error.
In our own previous work, Lotaru~\cite{baderLotaruLocallyEstimating2022}, we achieved a median prediction error between 14\% and 22\% for heterogeneous cluster infrastructures, while the prediction error for the best performing baseline yielded a median error of 31\%.
Further, our results showed that the prediction errors of tasks over a workflow are frequently distributed according to a long-tailed exponential distribution.

Accordingly to these observations, we introduce a prediction error noise to the scheduler input in our WorkflowSim fork.
Therefore, the predicted runtime $r_p$ is defined as ${r_p= r *( 1 \pm  \mathcal{N}(1, 0.5) * err})$ for a normal distribution and as $r_p$ as ${r_p= r *( 1 \pm  {Exp}(1) * err})$ for an exponential distribution where $r$ is the true runtime and $err$ the prediction error.

The respective prediction error is sampled from either a normal or an exponential distribution.
However, the actual task runtimes remain unchanged and are not influenced by the error.

Second, instead of relying on the runtime extrapolation by simply using the MIPS, our WorkflowSim fork looks up the real runtimes of a given task on a certain machine.
For our setup, we used 27 real instance types from AWS EC2, using the sizes \textit{large}, \textit{xlarge}, and \textit{2xlarge}, where applicable.
We run all evaluation workflows on these machines to gather detailed task runtimes.

Out of these 27 machines, we created 200 random heterogeneous clusters comprising of 40 nodes each.
All approaches run once on each of the 200 clusters.

\subsection{Evaluation Workflows}
\label{subsec:simulation-setup}

We selected three publicly available real-world Nextflow~\cite{nextflow} workflows from the popular nf-core~\cite{ewels2020nf} repository: \textit{Viralrecon}\footnote{github.com/nf-core/viralrecon, Accessed: June 2022} -- variant calling for viral samples, \textit{Eager} -- ancient DNA analysis\cite{peltzer2016eager}, and \textit{Chipseq}\footnote{github.com/nf-core/chipseq, Accessed: June 2022} -- peak-calling.
The workflows have different resource usage patterns, and different Directed-Acyclic-Graph (DAG) structures.

To feed our WorkflowSim extension with actual task-machine runtimes, we collect real trace information from real executions, which then can be extrapolated.
We run each workflow five times on the 27 instance types.
Further, we extended Nextflow in a way that it stores the traces in a WorkflowSim readable file.
The runtime is then extrapolated to simulate long-running workflows.

\subsection{Baselines}
We compare Reshi with three baselines, namely, Round-Robin, MinMin, and HEFT (Heterogeneous Earliest Finish Time).

The first baseline, Round-Robin, is a frequently used scheduling technique by resource managers, e.g., Kubernetes uses a round-robin like approach\cite{carrion2022kubernetes}.
MinMin is a popular dynamic job scheduling algorithm that orders the queue to be scheduled ascending by the task runtime and then selects the fastest machine~\cite{sharma2017comparative}.
HEFT~\cite{heft} is a static list scheduling algorithm that incorporates different task-machine runtimes, communication times, and the structure of the directed acyclic graph (DAG).
Except for Round-Robin, all baselines require a-priori knowledge about task-machine runtimes.

Since Reshi itself recommends machines and does not aim to schedule the tasks, e.g., prioritize the tasks by runtime or number of descendants, we show Reshi in combination with two simplified task prioritizing techniques.
Reshi-C compares the number of children tasks and prefers tasks with many children.
The second strategy, Reshi-M, orders all tasks by the average runtime from historical executions descending.

    \section{Evaluation Results}\label{sec:EXPERIMENTS}
    
\begin{table}[]
\centering
\caption{Workflow runtimes with different schedulers assuming a normally distributed runtime prediction error of 15\%.}
\begin{tabular}{|c|c|r|r|r|r|}
\hline
\multicolumn{1}{|l|}{}                                        & \multicolumn{1}{l|}{} & \textbf{Mean \%}    & \textbf{90p \%}      & \textbf{95p \%}      & \textbf{Max \%}      \\ \hline
                                                              & HEFT                  & 5.58           & 59.43           & 70.79           & 129.62          \\ \cline{2-6} 
                                                              & Reshi-C               & \textbf{0.00 } & \textbf{7.53 }  & \textbf{7.58 }  & \textbf{27.97 } \\ \cline{2-6} 
                                                              & Reshi-M               & 14.32          & 7.85            & 35.39           & 45.35           \\ \cline{2-6} 
                                                              & MinMin                & 8.71           & 22.44           & 28.70           & 42.76           \\ \cline{2-6} 
\multirow{-5}{*}{\textbf{Chipseq}}                            & RR                    & 69.41          & 116.35          & 119.21          & 122.61          \\ \hline
\rowcolor[HTML]{EFEFEF} 
\cellcolor[HTML]{EFEFEF}                                      & HEFT                  & 10.47          & 35.47           & 39.53           & 54.07           \\ \cline{2-6} 
\rowcolor[HTML]{EFEFEF} 
\cellcolor[HTML]{EFEFEF}                                      & Reshi-C               & 14.53          & 27.33           & 29.07           & 30.81           \\ \cline{2-6} 
\rowcolor[HTML]{EFEFEF} 
\cellcolor[HTML]{EFEFEF}                                      & Reshi-M               & \textbf{0.00 } & \textbf{6.40 }  & \textbf{6.40 }  & \textbf{16.28 } \\ \cline{2-6} 
\rowcolor[HTML]{EFEFEF} 
\cellcolor[HTML]{EFEFEF}                                      & MinMin                & 12.21          & 29.07           & 33.72           & 56.40           \\ \cline{2-6} 
\rowcolor[HTML]{EFEFEF} 
\multirow{-5}{*}{\cellcolor[HTML]{EFEFEF}\textbf{Eager}}      & RR                    & 52.33          & 99.42           & 119.19          & 119.77          \\ \hline

                                   & HEFT                  & 25.25          & 82.47           & 115.77          & 104.02          \\ \cline{2-6} 

                                    & Reshi-C               & 10.72          & 23.24           & 37.60           & 38.48           \\ \cline{2-6} 

                                    & Reshi-M               & \textbf{0.00 } & \textbf{15.30 } & \textbf{18.55 } & \textbf{29.08 } \\ \cline{2-6} 

                                    & MinMin                & 18.88          & 37.73           & 41.54           & 45.84           \\ \cline{2-6} 

\multirow{-5}{*}{\textbf{viralrecon}} & RR                    & 44.71          & 63.73           & 65.24           & 80.90           \\ \hline
\end{tabular}
\label{tab:normal_15_table}
\end{table}

\begin{figure*}
    \includegraphics[width=\textwidth]{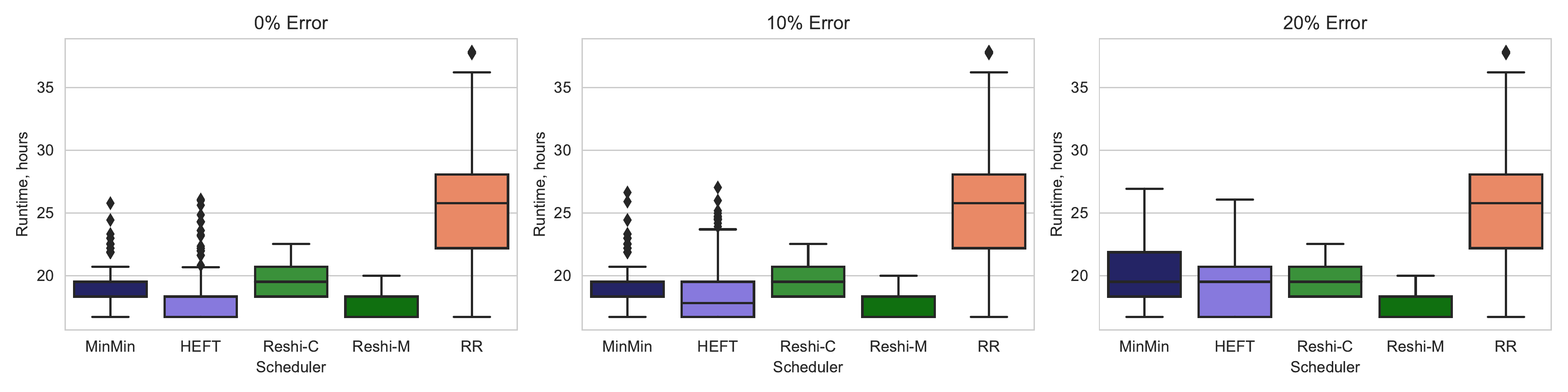}
    \caption{Makespans of the Eager workflow for different schedulers with normally distributed runtime prediction errors.} \label{fig:reshi_eager_normal}
\end{figure*}

We evaluate two different use-cases in our experiments using the same 200 clusters and three workflow setups each.
For both use-cases, we assume that the task runtime is predicted from historical execution traces~\cite{da2015online,nadeem2017modeling,pham2017predicting, hilman2018task}.
In the first use-case, we assume a prediction error that is normally distributed.
The second use-case incorporates an exponentially distributed prediction error.
We test both scenarios with different mean error rates that are based on reported research results, as elaborated in Section~\ref{subsec:workSim}.

Due to the high number of evaluation setups, the following sections report detailed results for the Eager workflow with different error assumptions and results for all workflows with a systematic over- or under-prediction of 15\%.

\subsection{Normally Distributed Error}

In our first scenario, we compare Reshi's resource allocation with Round-Robin and two state-of-the-art schedulers assuming a normally distributed prediction error.
Figure~\ref{fig:reshi_eager_normal} shows the makespan of the Eager workflow with different normally distributed runtime prediction errors.
\mbox{Reshi-M} and HEFT achieve the same median makespans assuming a totally accurate task runtime, i.e., no prediction error.
However, once the error increases, all schedulers except for Reshi-C, \mbox{Reshi-M}, and Round-Robin lead to higher makespans. 
With an error of 15\%, HEFT yields a 75th percentile that has a 9.81\% higher makespan compared to the 75th percentile of \mbox{Reshi-M}.
\mbox{Reshi-M's} percentile is always below the respective 75th percentile of the competitors.
For all the baseline approaches, except for Round-Robin, an increased error leads to longer workflow runtimes and more cases where outliers, i.e., workflow executions with longer makespans, can be detected.

\begin{figure*}[ht!]
    \includegraphics[width=\textwidth]{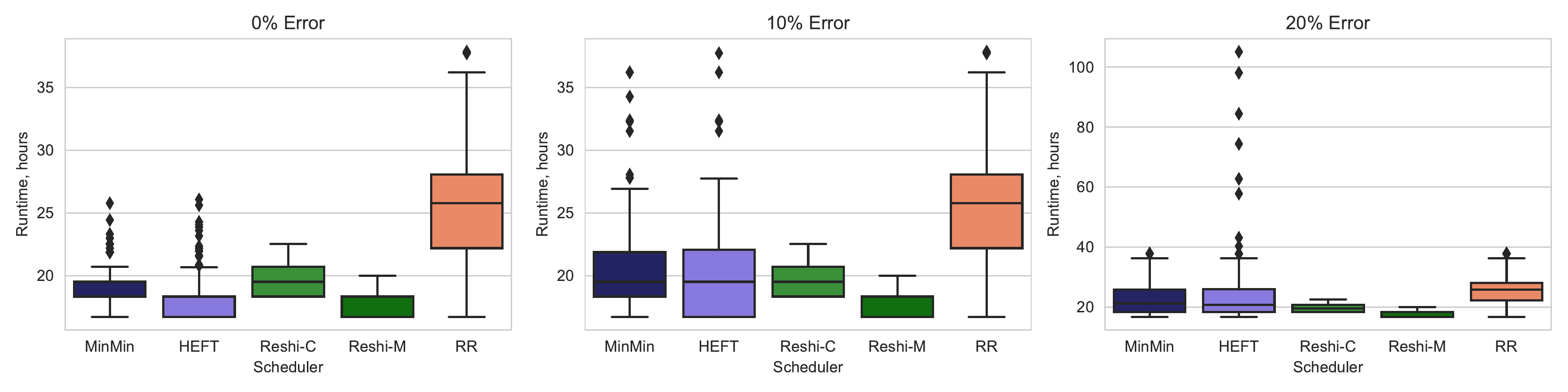}
    \caption{Makespans of the Eager workflow for different schedulers with exponentially distributed runtime prediction errors.} \label{fig:reshi_eager_exponential}
\end{figure*}
\begin{table}[]
\centering
\caption{Workflow runtimes with different schedulers assuming an exponentially distributed runtime prediction error of 15\%.}
\begin{tabular}{|l|r|r|r|r|r|}
\hline
\multicolumn{1}{|l|}{}                                        & \multicolumn{1}{l|}{} & \textbf{Mean \%}    & \textbf{90p \%}      & \textbf{95p \%}      & \textbf{Max \%}      \\ \hline
                                                              & HEFT                  & 21.86          & 76.56           & 92.49           & 229.37          \\ \cline{2-6} 
                                                              & Reshi-C               & \textbf{0.00 } & \textbf{4.98 }  & \textbf{5.03 }  & \textbf{24.94 } \\ \cline{2-6} 
                                                              & Reshi-M               & 11.61          & 31.63           & 32.18           & 41.90           \\ \cline{2-6} 
                                                              & MinMin                & 15.41          & 41.40           & 51.86           & 115.54          \\ \cline{2-6} 
\multirow{-5}{*}{\textbf{Chipseq}}                            & RR                    & 65.39          & 111.22          & 114.01          & 117.33          \\ \hline
\rowcolor[HTML]{EFEFEF} 
\cellcolor[HTML]{EFEFEF}                                      & HEFT                  & 23.84          & 54.65           & 87.79           & 119.77          \\ \cline{2-6} 
\rowcolor[HTML]{EFEFEF} 
\cellcolor[HTML]{EFEFEF}                                      & Reshi-C               & 14.53          & 27.33           & 29.07           & 30.81           \\ \cline{2-6} 
\rowcolor[HTML]{EFEFEF} 
\cellcolor[HTML]{EFEFEF}                                      & Reshi-M               & \textbf{0.00 } & \textbf{6.40 }  & \textbf{6.40 }  & \textbf{16.28 } \\ \cline{2-6} 
\rowcolor[HTML]{EFEFEF} 
\cellcolor[HTML]{EFEFEF}                                      & MinMin                & 29.07          & 83.72           & 99.42           & 119.77          \\ \cline{2-6} 
\rowcolor[HTML]{EFEFEF} 
\multirow{-5}{*}{\cellcolor[HTML]{EFEFEF}\textbf{Eager}}      & RR                    & 52.33          & 99.42           & 119.19          & 119.77          \\ \hline
                                   & HEFT                  & 35.83          & 73.37           & 86.69           & 198.87          \\ \cline{2-6} 
                                    & Reshi-C               & 10.53          & 23.24           & 37.60           & 38.48           \\ \cline{2-6} 
                                    & Reshi-M               & \textbf{0.00 } & \textbf{15.30 } & \textbf{18.55 } & \textbf{29.08 } \\ \cline{2-6} 
                                   & MinMin                & 28.17          & 57.91           & 62.18           & 77.78           \\ \cline{2-6} 

\multirow{-5}{*}{\textbf{viralrecon}} & RR                    & 44.71          & 63.73           & 65.24           & 80.90           \\ \hline
\end{tabular}
\label{tab:exp_15_table}
\end{table}

Table~\ref{tab:normal_15_table} summarizes all workflow makespans assuming a normally distributed task prediction error of 15\%.
For each workflow, we set the lowest mean value to 0 and depict the relative change according to that value.
One can see that two out of three times, Reshi-M achieves the lowest workflow runtimes and for the other workflow Reshi-C.
For Eager and viralrecon, Reshi-M achieves the best results.
This is, for example, due to the Eager workflow structure, where the maximum depth is two, and, therefore, strategies that prefer long graph dependencies, e.g., Reshi-C, yield to a higher makespan.
Reshi's 95th percentile is below the competitors' 90th percentile in all workflows.
Further, HEFT's maximum values are frequently more than two times higher compared to Reshi's maximum reported value.
For the Chipseq workflow, the maximum value is more than 4.5 times higher

\subsection{Exponentially Distributed Error}

In our second experiment we assume an exponentially distributed task runtime prediction error.
Again, Figure~\ref{fig:reshi_eager_exponential} shows the Eager workflow makespan for the different schedulers assuming various prediction error levels.
The maximum reported makespan of Reshi-M and Reshi-C is below the 75th percentile of HEFT.
Table~\ref{tab:exp_15_table} shows that Reshi's maximum value is always below the 90th percentile of the baseline approaches.
Further, the mean and the percentiles differences are much more considerable now.
Compared to the experiment with a normally distributed error from the section before, the baseline schedulers react stronger to an increase in the error rate, assuming an exponential error distribution.
Here, a higher error rate necessarily leads to higher makespans for the baselines, except for Round-Robin.
Since Reshi-C and \mbox{Reshi-M} do not rely on predicted runtimes or estimated resource usages, they constantly achieve the same results.

    \section{Conclusion}\label{sec:CONCLUSION}
    This paper presented an approach to dynamically map scientific workflow tasks onto heterogeneous infrastructures using our recommender systems.
Through the task-node ranks, Reshi does not rely on error-prone runtime estimates and has not to cope with data-dependent resource predictions.

Our experimental evaluation with three real-world Nextflow workflows from the popular nf-core framework shows that Reshi provides efficient task-machine allocations without requiring accurate task runtime estimates, while we show that in comparison the performance of state-of-the-art schedulers is highly dependent on accurate task-runtime predictions.

Pairing Reshi's recommendations with a simple scheduler, Reshi is able to outperform HEFT by a mean makespan reduction of 7.18\% for a normally distributed error of 15\% and a mean makespan reduction of 18.01\% for an exponentially distributed error of 15\%.
Further, the baseline schedulers yield significantly higher 90th percentile quarterlies and a significantly higher variance.

    \section*{Acknowledgments}
    \thanks{Funded by the Deutsche Forschungsgemeinschaft (DFG, German Research Foundation) as FONDA (Project 414984028, SFB 1404).}
    \bibliographystyle{IEEEtran}
    \bibliography{./references}

\end{document}